\documentclass[fleqn,usenatbib]{mnras}

\usepackage{newtxtext,newtxmath}

\usepackage[T1]{fontenc}
\usepackage{adjustbox}

\usepackage{graphicx}	

\title[Tidal tails in NGC~752]{Tidal tails in the disintegrating open cluster NGC~752}

\author[S. Bhattacharya et al.]{
Souradeep Bhattacharya,$^{1}$\thanks{E-mail: souradeep@iucaa.in}
Manan Agarwal,$^{2}$
Khushboo K. Rao$^{2}$
and Kaushar Vaidya$^{2}$
\\
$^{1}$Inter University Centre for Astronomy and Astrophysics, Ganeshkhind, Post Bag 4, Pune 411007, India\\
$^{2}$Department of Physics, Birla Institute of Technology and Science – Pilani, Rajasthan 333031, India
}

\date{Accepted 2021 May 12. Received 2021 May 10; in original form 2021 April 12}

\pubyear{2021}

\begin{document}
\label{firstpage}
\pagerange{\pageref{firstpage}--\pageref{lastpage}}
\maketitle

\begin{abstract}
We utilize the robust membership determination algorithm, ML-MOC, on the precise astrometric and deep photometric data from Gaia Early Data Release 3 within a region of radius 5$^{\circ}$ around the center of the intermediate-age galactic open cluster NGC~752 to identify its member stars. We report the discovery of the tidal tails of NGC~752, extending out to $\sim$35~pc on either side of its denser central region {and following the cluster orbit}. From comparison with PARSEC stellar isochrones, we obtain the mass function of the cluster with a slope, $\chi=-1.26\pm0.07$. The high negative value of $\chi$ is indicative of a disintegrating cluster undergoing mass-segregation. $\chi$ is more negative in the intra-tidal regions as compared to the outskirts of NGC~752. We estimate a present day mass of the cluster, M$\rm_{C}=297\pm10$ M$_{\sun}$. Through mass-loss due to stellar evolution and tidal interactions, we further estimate that NGC~752 has lost nearly 95.2-98.5\% of its initial mass, $\rm M_{i}~=~0.64~-2~\times~10^{4}~M_{\sun}$.
\end{abstract}




\section{Introduction}
\label{sect:intro}
Almost all stars ever formed were likely born in clustered environments \citep[see review by][and references therein]{Zwart10}. Stars inside a cluster are subjected to two-body relaxation which over time leads to more-massive stars sinking to the central regions of the cluster while less-massive stars occupy a larger volume, many of which gradually evaporate to supply the field population of a galaxy. This effect is called mass segregation and has been observed in both galactic globular \citep{hoerner57} and open \citep{Mathieu84} clusters, investigated both through the slope of the mass-function of the cluster members in the core and outskirts \citep[e.g.][]{Bhattacharya17b}, as well as through the radial distribution of massive objects, such as blue stragglers \citep[e.g.][]{Bhattacharya19,Vaidya20,Rao21}. Alongside this internal effect, Galactic potential perturbs star clusters leading to the formation of tidal structures. Such tidal structures are observed in both globular clusters (e.g. Palomar~5 - \citealt{Odenkirchen01}; NGC~5466 - \citealt{Belokurov06}) and open clusters (e.g. Berkeley~17 - \citealt{Chen04,Bhattacharya17b}; NGC~6791 - \citealt{Dalessandro15}). 

Galactic open clusters, in particular, are more abundant in the disc where they are additionally subjected to external gravitational effects such as interactions with giant molecular clouds and passage through spiral arms \citep{Spitzer58}. From an observational standpoint, the identification of open cluster morphology, despite the use of novel techniques \citep[e.g.][]{Bhattacharya17a}, has long since relied on stellar density determination against strong field contamination in photometric observations. Low-density features such as tidal tails have thus been particularly difficult to identify with little success \citep{Chen04,Dalessandro15}. Reliable astrometry (proper motion and parallax) from Gaia Data Release 2 \citep[Gaia DR2;][]{Gaia18} and recently Gaia Early Data Release 3 \citep[Gaia EDR3;][]{Gaia20} has allowed for improved field-star decontamination of many nearby clusters through novel membership determination techniques \citep{Agarwal21}. This has resulted in the recent discovery of a few open clusters {with extended stellar coronae \citep[e.g.][]{Carrera19,Meingast21} as well as those with} tidal features -- Hyades: \citet{Roser19}; Coma~Berenices: \citet{Tang19}; {Ruprecht 147: \citet{Yeh19};} Praesepe: \citet{Roser19b}; NGC~2506: \citet{Gao20}; M~67: \citet{Gao20b}; UBC~274: \citet{Castro20}; {Alpha Persei: \citet{Nikiforova20}}. 

NGC~752 (cluster center at RA=29.156 deg and DEC=37.809 deg; \citealt{Agarwal21}) is an intermediate age ($\sim1.5$ Gyr old; \citealt{Agueros18}), nearby ($\sim450$ pc) open cluster which was among the clusters whose membership determination was carried out by \citet{Agarwal21} using Gaia DR2 data with their novel membership determination algorithm, ML-MOC. Within their search radius of 2.5$^{\circ}$ around its center, \citet{Agarwal21} found hints of tidal tails in NGC~752. \citet{Hu21} also identify the elongated morphology of the peripheral regions of NGC~752 from the Gaia DR2 data using the members identified by \citet{cantat18}. 

In this paper, we report the discovery of the tidal tails of NGC~752 within a large spatial region around the cluster. The data and membership determination are presented in Section~\ref{sect:data}. The analysis of the mass function and tidal structure of NGC~752 is presented in Section~\ref{sect:analysis}. Finally, we discuss {the orbit of the cluster and its dissolution} in Section~\ref{sect:disc}.

\begin{figure*}
	\includegraphics[width=\textwidth]{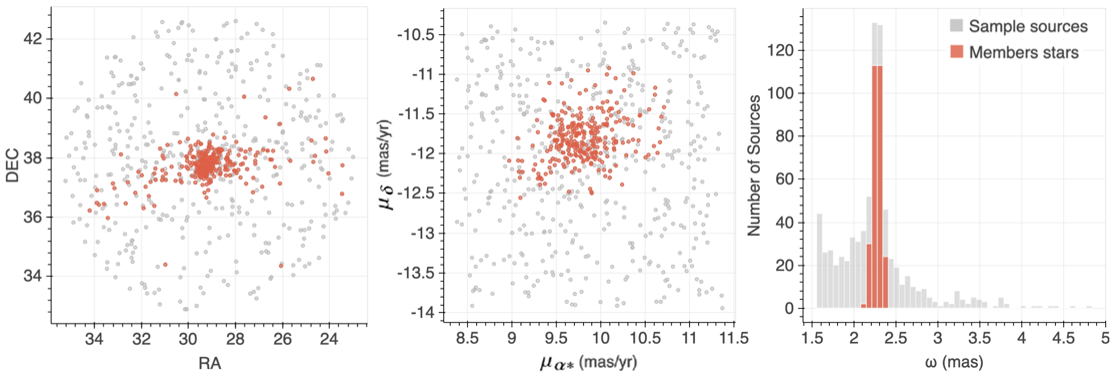}
    \caption{{The spatial, proper-motion and parallax distributions of the \textit{Sample sources} and the member stars identified by ML-MOC.}}
    \label{fig:mlmoc}
\end{figure*}

\begin{table*}
\caption{{Gaia EDR3 members identified in NGC~752 using ML-MOC. The full table will be available in the online version of the journal.}}
\centering
\adjustbox{max width=\textwidth}{
\begin{tabular}{ccccccccccc}
  \hline
    GaiaID & RA & DEC & $\omega$ & $\mu_{\alpha *}$ & $\mu_{\delta}$ & G & BP & RP & RV & $\rm p_{memb}$\\
     & deg & deg & mas & mas/yr & mas/yr & mag & mag & mag & km/s & \\    
  \hline
342914075859711744 & 29.23240573 & 37.79978973 & 2.288 & 9.803 & -11.878 & 11.1 & 11.3 & 10.75 & 6.91 & 0.98 \\
342912284857245952 & 29.15366539 & 37.75349579 & 2.272 & 9.892 & -12.043 & 13.64 & 14.04 & 13.07 & -- & 0.97 \\
342914556896034176 & 29.22108601 & 37.8692465 & 2.261 & 9.462 & -11.507 & 10.15 & 10.38 & 9.77 & 18.0 & 0.92 \\
342915411593428736 & 29.16345994 & 37.86139137 & 2.267 & 9.721 & -11.886 & 10.29 & 10.48 & 9.96 & 8.23 & 0.98 \\
342912254793586688 & 29.13259369 & 37.74945795 & 2.141 & 9.794 & -11.62 & 16.2 & 17.08 & 15.27 & -- & 0.68 \\
342913178210782080 & 29.11199587 & 37.79182891 & 2.231 & 9.854 & -12.06 & 17.83 & 19.17 & 16.72 & -- & 0.95 \\
342909845316110336 & 29.31580848 & 37.73702508 & 2.363 & 9.887 & -11.726 & 18.3 & 19.57 & 17.16 & -- & 0.91 \\
342920501130956800 & 29.26040912 & 37.88539331 & 2.235 & 9.815 & -11.984 & 11.21 & 11.43 & 10.84 & 5.1 & 0.97 \\
342920501130956928 & 29.26070082 & 37.88616266 & 2.219 & 10.006 & -11.428 & 13.9 & 14.2 & 13.24 & -- & 0.92 \\
342921257045010432 & 29.22252061 & 37.8972341 & 2.169 & 9.499 & -11.925 & 14.92 & 15.52 & 14.18 & -- & 0.9 \\
  \hline
\end{tabular}
\label{table:data}
}
\end{table*}

\begin{figure*}
	\includegraphics[width=\textwidth]{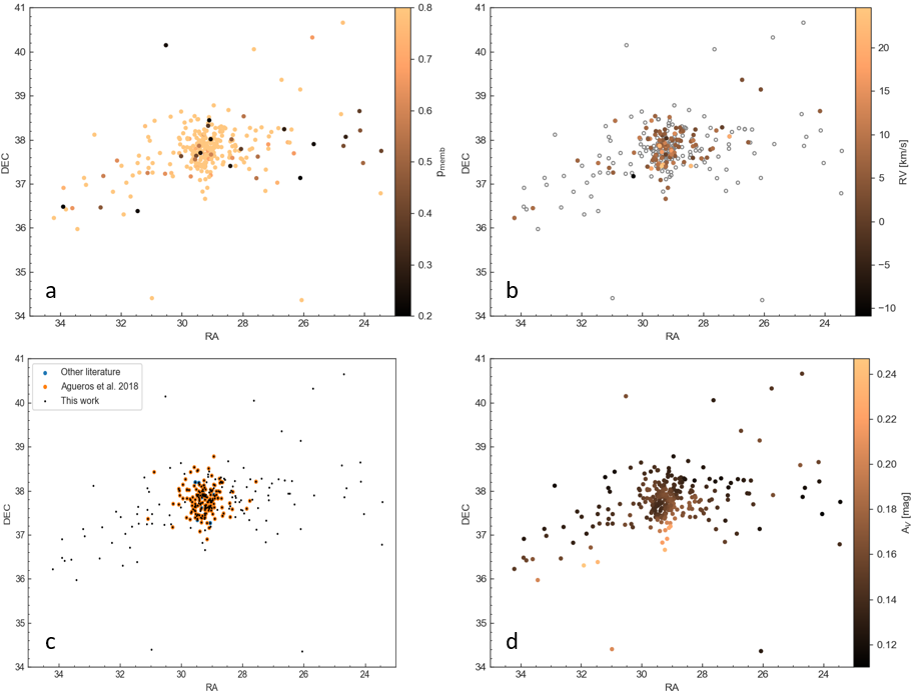}
    \caption{Spatial distribution of the identified members in NGC~752. (a) Coloured by their $\rm p_{memb}$ values. (b) Coloured by their RV values where available and open grey circles where unavailable.  (c) Matched previously-identified members. (d) Coloured by their $\rm A_{V}$ values.}
    \label{fig:spat}
\end{figure*}

\section{Data and membership selection}
\label{sect:data}
Gaia EDR3 provides the positions, trigonometric parallaxes ($\rm\omega$), and proper motions (PM) as well as photometry in three broad-band filters (G, BP, and RP) for almost 1.5 billion stars. It improves upon the Gaia DR2 with the proper motion uncertainty reduced on average by nearly a factor of two while parallax measurements are more accurate by $\sim$20\% \citep{Gaia20}. This allows for reliable photometry down to G$\sim$21 mag. Gaia EDR3 photometry is essentially complete down to G=17 mag although the completeness may be lower in regions of high stellar density such as those of globular clusters \citep{Fabricius20}. Radial velocity measurements in Gaia EDR3 are nearly the same as in Gaia DR2. Further details on Gaia EDR3 are available at \url{https://www.cosmos.esa.int/web/gaia/earlydr3}.

The membership selection in NGC~752 from the Gaia EDR3 data-set is carried out  in a region of radius 5$^{\circ}$ around its center. {This sample is termed \textit{All sources}.} We use ML-MOC \citep{Agarwal21} to identify cluster members using the proper motion and parallax information. While details of the membership selection are described in \citet{Agarwal21}, we briefly state it here. ML-MOC is based on the k-nearest neighbour algorithm \citep[kNN;][]{kNN} and the Gaussian mixture model \citep[GMM;][]{GMM}. It identifies cluster members in the PM--$\rm\omega$ parameter space, independent of the spatial density of the cluster, thereby allowing for the identification of faint extended spatial structures such as tidal tails. As a first step to the membership selection, kNN is utilised to remove the obvious field stars such that the remaining member candidates have more cluster members than field stars. {This is done by applying kNN to those stars within the central region of radius 1$^{\circ}$ to calculate a broad range of PM and $\omega$ values for the cluster members. For NGC~752, we find the range of PM in RA, $\mu_{\alpha *}=$ 8.372 -- 11.372 mas/yr, PM in DEC, $\mu_{\delta *}=$ -13.948 -- -10.348 mas/yr, and $\omega=$ 1.552 -- 5.199. The stars in \textit{All sources} which fall in the aforementioned parameter ranges are termed \textit{Sample sources}. Their spatial, PM and $\omega$ distributions are shown in Figure~\ref{fig:mlmoc}. } {For the second step, a three dimensional GMM is used in the PM--$\omega$ parameter space of the \textit{Sample sources}} to distinguish between the cluster and field members, also assigning a membership probability ($\rm p_{memb}$). Those sources having $\rm p_{memb}\geq0.6$ are considered as high probability members, also including all the radial velocity members. A small number of stars having $\rm 0.2 \leq p_{memb}\leq0.6$ are also considered as members if their $\rm\omega$ values lie within the range of $\rm\omega$ values specified by the members that have $\rm p_{memb}\geq0.8$. {This results in the identification of 282 members in NGC~752 whose spatial, PM and $\omega$ distributions are also shown in Figure~\ref{fig:mlmoc}.} Note the extended tails are clearly identified in the spatial distribution. {The identified members with their Gaia EDR3 IDs, astrometry, photometry, radial velocity (RV) and $\rm p_{memb}$ have been tabulated in Table~\ref{table:data}.}

In Figure~\ref{fig:spat}a, the marked $\rm p_{memb}$ values show that the {extended} tails are present for the high probability members showing their identification as being robust. The mean distance to the cluster is found to be 443.8 pc \citep[see][]{Agarwal21} consistent with literature values \citep{Agueros18}.  The cluster members with RV data from Gaia EDR3 are shown in Figure~\ref{fig:spat}b, having <RV>=7.88 km/s and $\rm\sigma_{RV}=5.03$ km/s. Few cluster members with RV data are also identified among the {extended} features. \citet{Agueros18} had identified 258 members in NGC~752 from photometry and PM from the Tycho-Gaia Astrometric Solution (TGAS) catalog \citep{Gaia16}. Furthermore, they had consolidated membership determinations from previous studies \citep[primarily][]{Daniel94,Mermilliod98} finding 32 literature members as non-members. 144 of the stars identified by \citep{Agueros18} as members have counterparts within 1$\arcsec$ of our identified members while 14 of the literature members classified as non-members also have counterparts within 1$\arcsec$ of our identified members. The culling of non-members in our membership sample as compared to those in the literature is testament to the improved accuracy of the Gaia EDR3 astrometry and our robust membership determination from ML-MOC. Figure~\ref{fig:spat}c shows that the literature members are concentrated in the central parts of NGC~752 as the {extended} tails were not surveyed in previous spatially-restrictive studies whose membership determinations were guided by cluster radial density.

\section{Analysis}
\label{sect:analysis}

\subsection{Reddening}
\label{sect:red}
We utilise the interstellar reddening, $E(B-V)$, map provided by \citet{SF11} available through the NASA/IPAC Infrared Science Archive\footnote{\url{https://irsa.ipac.caltech.edu/applications/DUST}} to obtain the line-of-sight extinction ($\rm A_{V} = 3.1 \times E(B-V)$; \citealt{cardelli89}) towards each identified cluster member in NGC~752. Figure~\ref{fig:spat}d shows the spatial distribution of the cluster members coloured by their $\rm A_{V}$. The mean extinction $\rm <A_{V}>=0.15$mag. The extinction for each cluster member is converted to the Gaia photometric system following the extinction law by \citet{Wang19}. There is indeed some evidence of differential reddening but only a small range of extinction values is covered. 

\begin{figure*}
	\includegraphics[width=\textwidth]{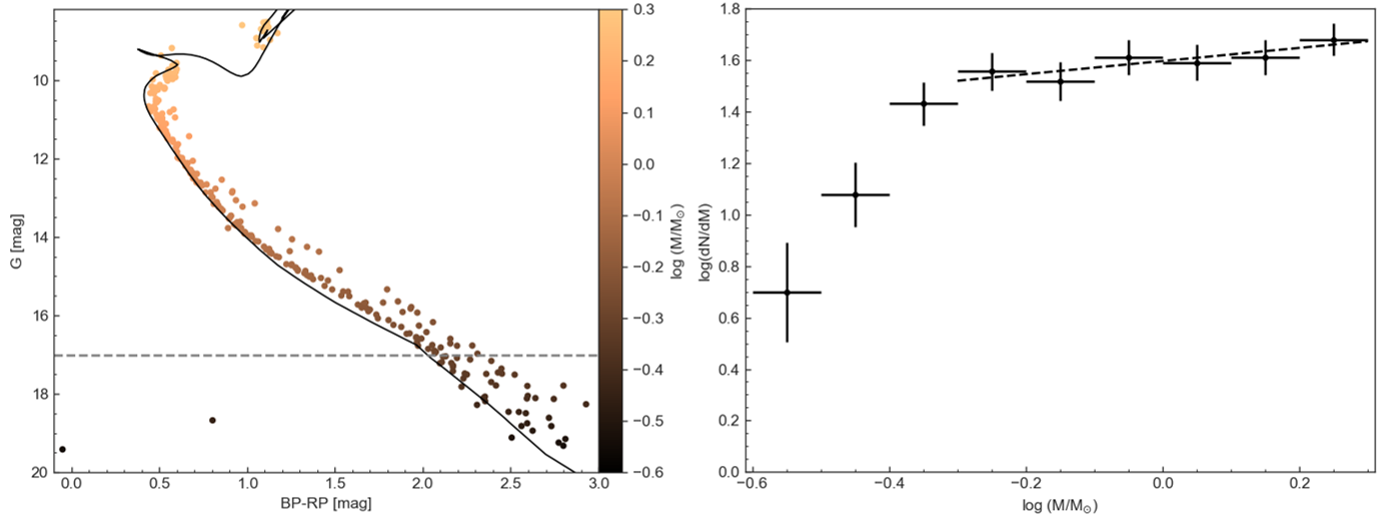}
    \caption{[Left] CMD of all the identified members in NGC~752, coloured by their stellar masses estimated from the PARSEC isochrone (black). The dashed grey line shows the $\sim$100\% completeness limit of the Gaia EDR3 data. [Right] The Mass function derived from the PARSEC iscochrones for the entire cluster. The dashed black line shows the fitted mass function where the Gaia EDR3 data is almost complete. The Poissonian uncertainties have been marked.}
    \label{fig:cmd}
\end{figure*}

\subsection{Color Magnitude Diagram and Mass Function}
\label{sect:cmd}
Correcting for extinction and reddening for each individual cluster member, we obtain the de-reddened color magnitude diagram (CMD) of the identified members shown in the left panel of Figure~\ref{fig:cmd}. The PARSEC stellar evolution isochrone\footnote{\url{http://stev.oapd.inaf.it/cgi-bin/cmd}} \citep{Bressan12,Chen14} for the Gaia filters has been plotted for the obtained distance corresponding to an age, t~=~1.5 Gyr and $\rm [M/H] = -0.04$, in line with fitted literature values of age (t~$=1.45\pm0.05$ Gyr) from multi-wavelength photometry by \citet{Twarog15} and metallicity ($\rm <[M/H]>=-0.07\pm0.04$) from spectroscopy of the red giant stars by \citet{Topcu15}. 

\begin{figure}
	\includegraphics[width=\columnwidth]{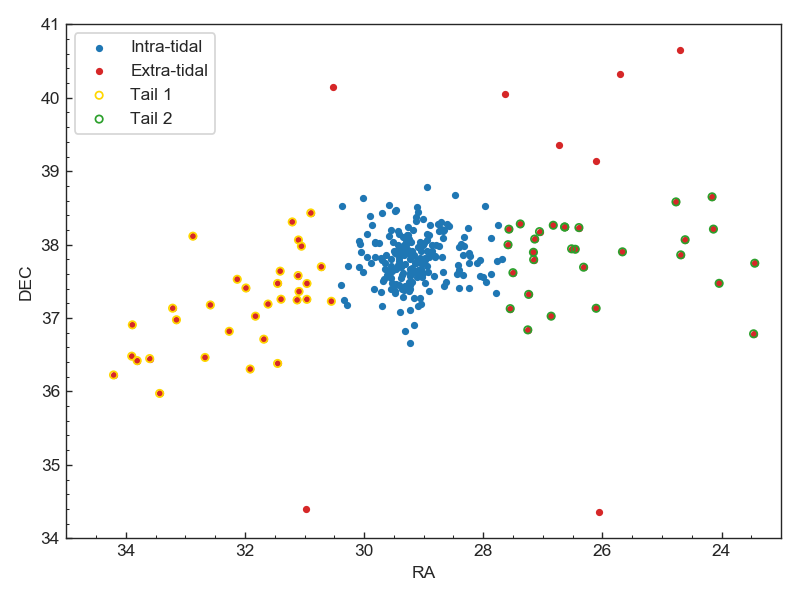}
    \caption{Spatial distribution of the identified members in NGC~752 with the intra-tidal (blue), extra-tidal (red), Tail~1 (encircled in yellow) and Tail~2 (encircled in green) members marked. }
    \label{fig:spat_tidal}
\end{figure}

Stellar mass of each star was estimated from the PARSEC isochrone and is shown in the left panel of Figure~\ref{fig:cmd}. The total observed mass of the cluster members is 284 M$_{\sun}$. The mass function of all the identified members in NGC~752 is shown in the right panel of Figure~\ref{fig:cmd}. The mass function is fitted above the stellar mass value ($\approx0.5$ M$_{\sun}$) corresponding to the Gaia EDR3 completeness limit of G=17 mag. Using the relation log(dN/dM)~=~-(1~+~$\chi$)~$\times$~log(M)~+~constant, where dN represents the number of stars in a mass bin dM with central mass M, we obtain the slope of the mass function for NGC~752, $\chi=-1.26\pm0.07$. Such a negative value of $\chi$, as compared to its value of $\chi=1.37$ as derived by \citep{salpeter55} for the solar neighbourhood, is indicative of a dissolving cluster that has undergone significant mass segregation. Extrapolating the fitted mass function to the stellar hydrogen burning limit of 0.08 M$_{\sun}$, we obtain the total mass of the cluster, M$\rm_{C}=297\pm10$ M$_{\sun}$. 

\begin{figure*}
	\includegraphics[width=0.9\textwidth]{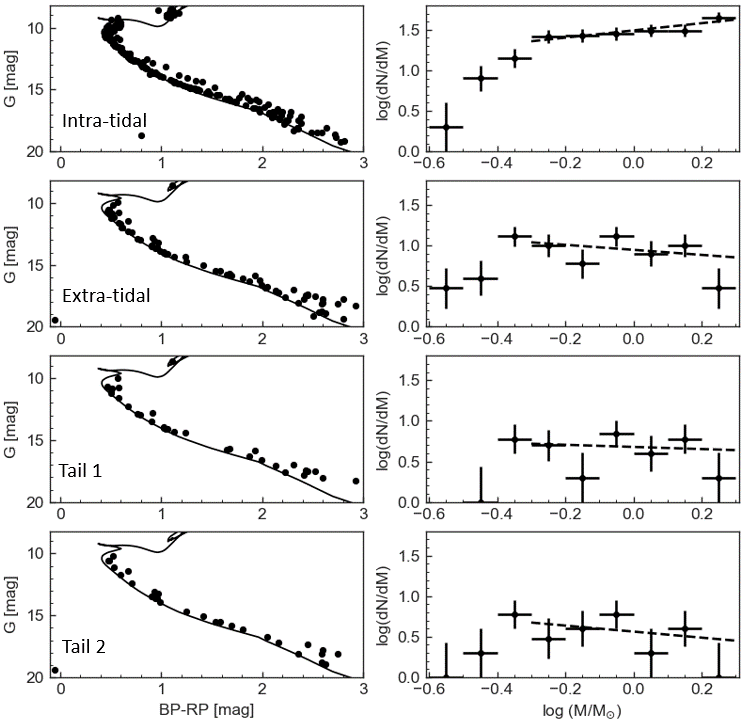}
    \caption{[Left] De-reddened CMD of all the identified members in NGC~752 in the four different regions. The black line shows the PARSEC isochrone corresponding to the age and metallicity of NGC~752. [Right] The Mass function derived from the PARSEC iscochrones for each region. The dashed black line shows the fitted mass function where the Gaia EDR3 data is almost complete. The Poissonian uncertainties have been marked.}
    \label{fig:mf_tidal}
\end{figure*}

\subsection{Tidal Structure}
\label{sect:tidal}
While extended tails are clearly visible in the spatial distribution of NGC~752 (Figure~\ref{fig:spat}), we need to estimate the tidal radius (r$\rm_{t}$) of the cluster in order to ascertain the {extra-tidal} nature of these tails. As estimating the tidal radius from fitting the King function \citep{King66} to the radial density of the cluster \citep[as carried out by][]{Agarwal21} is not appropriate for a dissolving cluster with extended structures, we calculate r$\rm_{t}$ following \citet{Pinfield98} as:
\begin{equation}
    $$\rm r_{t}^{3} = \rm \frac{G M_{C}}{2(A-B)^{2}}$$
\end{equation}
where G is the gravitational constant, and A and B are the
Oort constants (A$=15.3\pm0.4$ km/s/kpc, B$=-11.9\pm0.4$ km/s/kpc; \citealt{Bovy17}). We calculate r$\rm_{t}=1.2123^{\circ}=9.52$~pc at the distance of NGC 752. The identified members within and beyond r$\rm_{t}$ are classified as intra-tidal and extra-tidal members respectively  (Figure~\ref{fig:spat_tidal}). Amongst the extra-tidal members, we further note those stars belonging to either of the two tails (1 \& 2; see Figure~\ref{fig:spat_tidal}). At the distance of NGC~752, tail 1 and 2 extend out to $\sim$34 and $\sim$36.5 pc respectively from the center of the cluster.

We further obtain the de-reddened CMDs for the four sub-regions of NGC~752 (intra-tidal, extra-tidal, Tail 1 and Tail 2) as shown in the left panel of Figure~\ref{fig:mf_tidal}. The mass functions for each of the sub-regions are obtained and fitted as shown in the right panel of Figure~\ref{fig:mf_tidal}. The fitted slopes of the mass functions are $\chi=-1.44\pm0.12$, $-0.69\pm0.49$, $-0.87\pm0.57$ and $-0.63\pm0.64$ for the intra-tidal, extra-tidal, Tail 1 and Tail 2 respectively. The mass function slope is most negative in the intra-tidal region, as expected in case of mass segregation, while it is relatively less-steep outside r$\rm_{t}$ despite the large uncertainty. 

\section{Discussion}
\label{sect:disc}

\begin{figure*}
	\includegraphics[width=\textwidth]{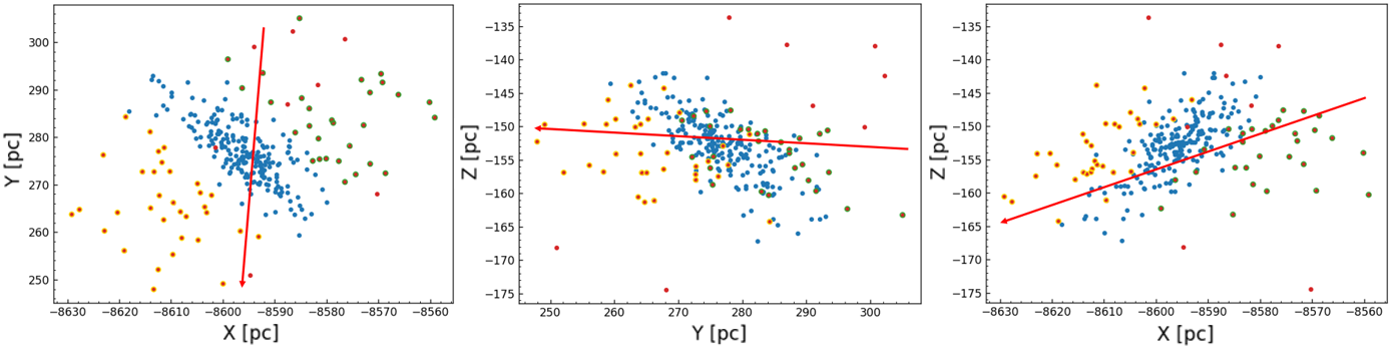}
    \caption{{Tri-dimensional projections in the galactocentric coordinate system of the identified members in NGC~752 with the intra-tidal (blue), extra-tidal (red), Tail~1 (encircled in yellow) and Tail~2 (encircled in green) members marked. The red arrow follows the cluster orbit. }}
    \label{fig:orbit}
\end{figure*}

\subsection{Cluster orbit}
\label{sect:orbit}

{To further validate the tidal origin of the extended tails, we explore their position with respect to the cluster orbit in a galactocentric coordinate system, similar to that carried out by \citet{Tang19} for Coma Berenices. The Galactocentric coordinates have a positive x direction pointing from the position of the Sun projected to the Galactic midplane to the Galactic center (GC); the y-axis points toward \textit{l}=90$^{\circ}$ and the z-axis roughly points toward \textit{b}=90$^{\circ}$.  Distance to the GC is 8.3 kpc \citep{Reid04}. Figure~\ref{fig:orbit} shows the X-Y (left), Y-Z (middle) and X-Z (right) projections of the cluster members.} 

{We compute the average position of all member stars as the cluster mean position giving X=-8595.62 pc, Y=276.26 pc, Z=-152.36 pc, and its mean space velocity (for only those sources having RV measurements) as U=-5.13 km/s, V=226.44 km/s, W=-12.03 km/s. The aforementioned are utilised to calculate the cluster orbit using the python package galpy \citep{Bovy15} which accounts for solar motion and includes the Milky Way potential model \textit{MWPotential2014} (A fitted Galactic potential model incorporating the bulge, disc and halo). Orbits are integrated 100 Myr backward and 100 Myr forward in time.} 

{Figure~\ref{fig:orbit} shows the orbit motion of NGC~752 in the three galactocentric projections. The distribution of the extended tails along the cluster orbit is most prominent in the Y-Z plane, similar to that found by \citet{Roser19}, \citet{Tang19} and \citet{Yeh19} for their studied clusters. This is a testament to the tidal nature of the extended tails of NGC~752 with Tails 1 \& 2 being the leading and trailing tails respectively.} Through reliable membership determination up to large radii in conjunction with reliable astrometry from Gaia EDR3, NGC~752 is now added to a growing list of dissolving open clusters exhibiting tidal features.

\subsection{Initial cluster mass}
\label{sect:inimass}
The mass-segregation of NGC~752, evident from the negative value of its mass function slope, and extended tidal features mark NGC~752 as a telltale sign of its dissolution. Environmental effects like tidal shocks due to close interactions with giant molecular clouds, spiral arms, the Galactic disc and, in
general, interactions with the Galactic tidal field or internal dynamical effects like two-body relaxation lead to mass-loss in an open cluster \citep{Dalessandro15}. 

\citet{Lamers05} provide an analytical prescription to estimate the mass lost by an open cluster due to the Galactic tidal field. The initial mass ($\rm M_{i}$) is estimated with the following relation:
\begin{equation}
    $$\rm M_{i} = \rm \bigg[\bigg(\frac{M_{C}}{M_{\sun}}\bigg) ^{\gamma} + \frac{\gamma t}{t_{0}} \bigg]^{\frac{1}{\gamma}} [1-q_{ev}(t)]^{-1}$$
    \label{eq:mini}
\end{equation}
where t$\rm_{0}$ is the dissolution time-scale parameter. By comparing the distribution of mass and age of OCs in the solar neighbourhood with theoretical predictions, \citet{Lamers05} obtained t$\rm_{0}=3.3_{-1.0}^{1.4}$ Myr. $\rm \gamma$ is a dimensionless index which depends on the cluster initial
density distribution. We adopt $\rm \gamma=0.62$ which is the typical value for open clusters \citep{Lamers05, Dalessandro15}. The function q$\rm_{ev}$(t) describes the mass-loss due to stellar evolution and can be approximated by the following analytical formula:
\begin{equation}
    $$\rm q_{ev}(t) = \rm (log_{10}(t)-a)^{b}+c$$
    \label{eq:qev}
\end{equation}
where a = $7.03$, b = $0.26$ and c = $-1.80$ for the metallicity of NGC 752 \citep[see Table 1 in][]{Lamers05}. Plugging the values in to Equations~\ref{eq:mini}~\&~~\ref{eq:qev}, we find $\rm M_{i}=11087_{-4666}^{8922}$, i.e,  $\rm M_{i}~=~0.64~-2~\times~10^{4}~M_{\sun}$. Thus, NGC~752 is a descendent of a Young Massive Cluster (which have stellar masses $\gtrsim 10^{4} \rm M_{\sun}$) such as those observed in the Milky Way and other galaxies \citep[see review by][]{Zwart10} that has lost 95.2--98.5\% of its mass to the Galactic field. 

\section*{Acknowledgements}
{This work presents results from the European Space Agency (ESA) space mission Gaia. Gaia data are being processed by the Gaia Data Processing and Analysis Consortium (DPAC). Funding for the DPAC is provided by national institutions, in particular the institutions participating in the Gaia MultiLateral Agreement (MLA). We thank the anonymous referee for their comments.} This research made use of Astropy-- a community-developed core Python package for Astronomy \citep{Astropy13}, SciPy \citep{scipy}, NumPy \citep{numpy} and Matplotlib \citep{matplotlib}. This research also made use of NASA’s Astrophysics Data System (ADS\footnote{\url{https://ui.adsabs.harvard.edu}}).

\section*{Data Availability}
The data underlying this article are publicly available at \url{https://archives.esac.esa.int/gaia}. {The 282 identified members of NGC~752 along with their Gaia EDR3 identifiers and photometric and astrometric data have been presented in Table~\ref{table:data}}.



\bibliographystyle{mnras}
\bibliography{ref_oc} 







\bsp	
\label{lastpage}
\end{document}